\documentclass[pra,twocolumn,superscriptaddress,preprintnumbers,showpacs]{revtex4}
\usepackage{amsmath,amssymb,bm,graphicx}

\renewcommand\d{\partial}
\newcommand\grad{\bm{\nabla}}
\newcommand\+{\dagger}
\renewcommand\r{{\bm{r}}}
\newcommand\x{{\bm{x}}}
\newcommand\p{{\bm{p}}}
\newcommand\eff{\mathrm{eff}}
\newcommand\0{\bm{0}}

\begin{document}
\preprint{MIT-CTP 4026}

\title{Confinement-induced Efimov resonances in Fermi-Fermi mixtures}
\author{Yusuke~Nishida}
\affiliation{Center for Theoretical Physics,
Massachusetts Institute of Technology, Cambridge, Massachusetts 02139, USA}
\author{Shina~Tan}
\affiliation{Department of Physics,
Yale University, New Haven, Connecticut 06520, USA}

\begin{abstract}
 A Fermi-Fermi mixture of ${}^{40}$K and ${}^6$Li does not exhibit the
 Efimov effect in a free space, but the Efimov effect can be induced by
 confining only ${}^{40}$K in one dimension.  Here the Efimov's
 three-body parameter is controlled by the confinement length.  We show
 that the three-body recombination rate in such a system in the dilute
 limit has a characteristic logarithmic-periodic dependence on the
 effective scattering length with the scaling factor 22.0 and can be
 expressed by formulas similar to those for identical bosons in three
 dimensions.  The ultracold mixture of ${}^{40}$K and ${}^6$Li in the
 one-dimensional$-$three-dimensional mixed dimensions is thus a
 promising candidate to observe the Efimov physics in fermions.
\end{abstract}

\date{March 2009}


\maketitle

\section{Introduction}
Recent realization of an ultracold Fermi-Fermi mixture of ${}^{40}$K and
${}^6$Li with interspecies Feshbach resonances opens up new research
directions in cold atomic
physics~\cite{Taglieber:2008,Wille:2008,Voigt:2008}.  Such examples
include the creation of the Bose-Einstein condensate of heteronuclear
molecules and the investigation of the effect of mass difference on the
superfluidity.  More importantly, the two-species mixture offers the
possibility of species-selective confinement potentials, which provides
novel subjects such as Fermi gases imbalanced in terms of the
dimensionality of space~\cite{Nishida:2008kr,Nishida:2008gk}.

It has been shown in Ref.~\cite{Nishida:2008kr} that when ${}^{40}$K is
confined in one dimension or two dimensions with ${}^6$Li being in three
dimensions, the ${}^{40}$K-${}^6$Li mixture with a resonant interspecies
interaction exhibits the Efimov effect characterized by an infinite
series of shallow three-body bound states (trimers) composed of two
heavy and one light fermions.  Their binding energies are given by
\begin{equation}\label{eq:efimov}
 E_3^{(n)} \to - e^{-2\pi n/s_0}\frac{\hbar^2\kappa_*^2}{2m_\mathrm{KLi}}
  \qquad\text{for}\qquad n\to\infty,
\end{equation}
with
$m_\mathrm{KLi}=m_\mathrm{K}m_\mathrm{Li}/(m_\mathrm{K}+m_\mathrm{Li})$
being the reduced mass and $s_0=1.02$ 
in the one-dimensional$-$three-dimensional (1D-3D) mixed dimensions and
$s_0=0.260$ 
in the two-dimensional$-$three-dimensional (2D-3D) mixed
dimensions~\cite{Nishida:2008kr}.  $\kappa_*$ is the so-called Efimov
parameter defined up to multiplicative factors of $e^{\pi/s_0}$ and will
be calculated in this Rapid Communication.  The emergence of such Efimov
trimers in the ${}^{40}$K-${}^6$Li mixture is remarkable because they
are absent in a free space but induced by confining ${}^{40}$K in lower
dimensions.  An alternative way to realize the Efimov effect using the
${}^{40}$K-${}^6$Li mixture would be to apply an optical lattice to
${}^{40}$K to increase its effective mass by more than a factor of
2~\cite{Petrov:2007}.

In this Rapid Communication, we will show that the Efimov effect in the
${}^{40}$K-${}^6$Li mixture when ${}^{40}$K is confined in 1D, is
experimentally observable through the three-body recombination rate
which has a characteristic logarithmic-periodic behavior as a function
of the effective scattering length.  In particular, the three-body
recombination rate is found to exhibit resonant peaks that can be
explained by {\em confinement-induced Efimov resonances\/}.  We note
that the three-body recombination rate has been successfully employed to
obtain evidences for the Efimov trimers in a Bose gas of
${}^{133}$Cs~\cite{Kraemer:2006,Knoop:2009}, a Bose-Bose mixture of
${}^{87}$Rb and ${}^{41}$K~\cite{Barontini:2009}, and a three-component
Fermi gas of ${}^6$Li~\cite{Ottenstein:2008,Huckans:2009}.

\section{Efimov parameter in the Born-Oppenheimer approximation}
Before developing an exact analysis of the three-body recombination
rate, it is worthwhile to demonstrate how the Efimov effect is realized
by confining ${}^{40}$K in 1D.  For generality, we shall consider a
three-body problem of two $A$ atoms and one $B$ atom with the resonant
interspecies interaction in which a two-dimensional harmonic potential
is applied to only $A$ atoms.  When the $A$ atoms are much heavier than
the $B$ atom $m_A\gg m_B$, one can first solve the Schr\"odinger
equation for the $B$ atom with fixed positions of the $A$ atoms, which
generates the following effective potential between two $A$ atoms:
$V(\r)=-\hbar^2c^2/(2m_Br^2)$, with $c=0.567$. 
Then, the relative motion of $A$ atoms is governed by the Schr\"odinger
equation (here and below $\hbar=1$)
\begin{equation}\label{eq:schrodinger}
 \left[-\frac{\grad_{\!\r}^2}{m_A}
  +\frac14m_A\omega^2\x^2+V(\r)\right]\Psi(\r)
 = \left(E_3+\omega\right)\Psi(\r),
\end{equation}
where $\omega$ is the oscillator frequency and $\r=(z,\x)$ with
$\x=(x,y)$ is relative coordinates between two $A$ atoms.  Fermi
statistics of $A$ atoms implies $\Psi(-\r)=-\Psi(\r)$.

In a free space $\omega=0$, it is known that the mass ratio
$m_A/m_B=6.67$ for the ${}^{40}$K-${}^6$Li mixture is too small to
form Efimov trimers~\cite{Petrov:2003}.  However, the confinement
potential term in Eq.~(\ref{eq:schrodinger}) makes it possible by
effectively reducing the dimensionality of $A$ atoms.  When
$m_A/m_B>1/(2c^2)=1.55$, 
bound-state wave functions with $E_3<0$ behave at long distance
$|z|\gg l$ as
\begin{equation}\label{eq:asymptotic}
 \Psi(\r) \to e^{-|\x|^2/(4l^2)}\frac{|z|^{3/2}}{z}
  K_{i\nu}\bigl(\sqrt{m_A|E_3|}|z|\bigr).
\end{equation}
Here $\nu\equiv\sqrt{\frac{c^2m_A}{2m_B}-\frac14}$ and
$l\equiv\frac1{\sqrt{m_A\omega}}$ is the confinement length.  For shallow
bound states $E_3\to-0$, the Bessel function in
Eq.~(\ref{eq:asymptotic}) oscillates as
$K_{i\nu}\bigl(\sqrt{m_A|E_3|}|z|\bigr)\propto\sin\bigl\{\nu\ln\bigl(\sqrt{m_A|E_3|}|z|/2\bigr)-\arg[\Gamma(1+i\nu)]\bigr\}$,
and their binding energies can be determined by matching this asymptotic
behavior with the numerical solution of Eq.~(\ref{eq:schrodinger}) for
$E_3=0$.  The oscillating asymptotic behavior implies that there exists
an infinite number of bound states with two successive binding energies
separated by a factor of $e^{2\pi/\nu}$.  In particular, in the case
of ${}^{40}$K-${}^6$Li mixture with $m_A/m_B=6.67$, we find
$E_3^{(n)}\to-14.0\,e^{-2\pi n/\nu}/(m_Al^2)$ 
for $n\to\infty$, from which we obtain $s_0\approx\nu$ and the Efimov
parameter $\kappa_*$ in Eq.~(\ref{eq:efimov}) as
$\kappa_*\approx1.91/l$. 

One should bear in mind that those numbers may not be accurate because
of the Born-Oppenheimer approximation we employed~\cite{error}.
However, the analysis presented here reveals the remarkable qualitative
aspect of the Efimov effect induced by the confinement potential: the
Efimov parameter is determined by the confinement length, and therefore
it is tunable by an external optical trap to a certain extent.  This is
in sharp contrast to other systems in a free space where Efimov
parameters are determined by short-range physics that is in general
difficult to control.

If the confinement length $l$ is much smaller than mean interatomic
distances and the thermal de Broglie wavelength of the system at finite
densities and temperature, one can consider $A$ atoms to be fixed on the
1D line neglecting their motion in the confinement direction.
Consequently, the resulting system becomes a two-species Fermi gas in
the 1D-3D mixed dimensions~\cite{Nishida:2008kr}.  However, when the
Efimov effect is present, the confinement length scale can not be
removed from the problem completely but appears as the Efimov parameter
$\kappa_*$ in the three-body scattering problem as we will see below.

\section{Effective field theory approach}
In order to develop a model-independent analysis of the Efimov effect in
our system, it is useful to adopt an effective field theory approach,
which has been a powerful method to study the Efimov physics in
identical bosons~\cite{Bedaque:1998kg}.  The two-species fermions in the
1D-3D mixed dimensions is universally described by the
action~\cite{Nishida:2008gk}
\begin{align}
 S &= \int\!dt\!\int\!dz
 \left[\psi_A^\+\left(i\d_t+\frac{\nabla_{\!z}^2}{2m_A}\right)\psi_A
 + g_0\psi_A^\+\psi_B^\+\psi_B\psi_A\right] \notag\\
 &\quad + \int\!dt\!\int\!dz\,d\x\,\psi_B^\+
 \left(i\d_t+\frac{\nabla_{\!z}^2+\grad_{\!\x}^2}{2m_B}\right)\psi_B.
\end{align}
Here $\psi_A(t,z)$ and $\psi_B(t,z,\x)$ are fermionic fields for $A$
atoms in 1D and $B$ atoms in 3D, respectively.  Their bare propagators
in the momentum space are given by $i/\left[p_0-p_z^2/2m_A+i0^+\right]$
for $A$ atoms and  $i/\left[p_0-(p_z^2+\p^2)/2m_B+i0^+\right]$ for $B$
atoms with $\p=(p_x,p_y)$.  The interspecies short-range interaction
takes place only on the 1D line located at $\x=\0$ and thus the
interaction term should be read as
$g_0\psi_A^\+(t,z)\psi_B^\+(t,z,\0)\psi_B(t,z,\0)\psi_A(t,z)$.

\begin{figure}[tp]
 \includegraphics[width=0.48\textwidth,clip]{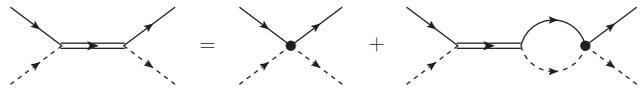}
 \caption{Two-body scattering of $A$ (solid line) and $B$ (dotted line)
 atoms.  The double line represents the scattering amplitude
 $i\mathcal{A}_2$.  \label{fig:2-body}}
\end{figure}

The two-body scattering of $A$ and $B$ atoms is depicted in
Fig.~\ref{fig:2-body}, and its amplitude $\mathcal{A}_2$ is given
by~\cite{Nishida:2008gk}
\begin{equation}
\mathcal{A}_2(p_0,p_z) = \frac{2\pi}{m_B}
  \frac1{\frac1{a_\eff} - \sqrt{\frac{m_{AB}}{M}p_z^2-2m_{AB}p_0-i0^+}},
\end{equation}
where $M=m_A+m_B$ is the total mass.  Here the effective scattering
length $a_\eff$ is introduced, which is related to the bare coupling
$g_0$ and the ultraviolet cutoff $\Lambda$ by
$\frac1{g_0}-\frac{\sqrt{m_Bm_{AB}}}{2\pi}\Lambda=-\frac{m_B}{2\pi a_\eff}$.
$a_\eff\to-0$ corresponds to the weak attraction and $a_\eff\to+0$
corresponds to the strong attraction between $A$ and $B$ atoms.  When
$a_\eff>0$, there exists a shallow two-body bound state (dimer) whose
binding energy $E_2=-\frac1{2m_{AB}a_\eff^2}$ is obtained as a pole of
the scattering amplitude in the center-of-mass frame:
$\mathcal{A}_2(E_2,0)^{-1}=0$.
The dependence of $a_\eff$ on the scattering length $a$ in a free space,
which is arbitrarily tunable by means of the interspecies Feshbach
resonance, was determined when the $A$ atom is confined in 1D by a
harmonic potential~\cite{Nishida:2008kr}.

\begin{figure}[tp]
 \includegraphics[width=0.48\textwidth,clip]{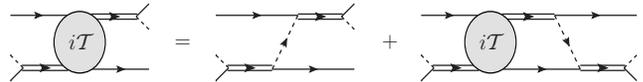}
 \caption{Three-body scattering of two $A$ and one $B$ atoms.
 \label{fig:3-body}}
\end{figure}

We now proceed to the three-body scattering of two $A$ atoms and one $B$
atom and show the existence of the Efimov trimers.  All the relevant
diagrams can be summed by solving the integral equation for
$\mathcal{T}(E;p_z,q_z)$, which is depicted in Fig.~\ref{fig:3-body}.
Here $E$ is the total energy in the center-of-mass frame and $p_z$
($q_z$) is the momentum of the incoming (outgoing) $A$ atom.
$\mathcal{T}$ has a property
$\mathcal{T}(E;p_z,q_z)=\mathcal{T}(E;-p_z,-q_z)$ and can be decomposed
into even- and odd-parity parts;
$\mathcal{T}_\mathrm{e(o)}(E;p_z,q_z)\equiv[\mathcal{T}(E;p_z,q_z)\pm\mathcal{T}(E;p_z,-q_z)]/2$.
The Efimov effect arises in the odd-parity channel
$\mathcal{T}_\mathrm{o}$, which satisfies an integral equation
\begin{equation}\label{eq:odd-parity}
 \begin{split}
  & \mathcal{T}_\mathrm{o}(E;p_z,q_z)
  = \frac{m_B}{4\pi}K(E+i0^+;p_z,q_z) \\
  &\quad + \int_0^\Lambda\!\frac{dk_z}{2\pi}
  \frac{\mathcal{T}_\mathrm{o}(E;p_z,k_z)\,K(E+i0^+;k_z,q_z)}
  {\sqrt{\frac{m_B+m_{AB}}{M}k_z^2-2m_{AB}E-i0^+}-\frac1{a_\eff}}
 \end{split}
\end{equation}
with
\begin{equation}
 K(E;p_z,q_z)
  = \ln\!\left(\frac{p_z^2+q_z^2+\frac{2m_A}{M}p_zq_z-2m_{AB}E}
	  {p_z^2+q_z^2-\frac{2m_A}{M}p_zq_z-2m_{AB}E}\right).
\end{equation}

When $m_A/m_B>2.06$, the integration over $k_z$ has to be cut off by
$\Lambda\sim l^{-1}$ and the limit $\Lambda\to\infty$ can not be taken.
Instead, the dependence on the arbitrary cutoff $\Lambda$ can be
eliminated by relating it to the physical parameter $\kappa_*$ defined
in Eq.~(\ref{eq:efimov}).  The spectrum of three-body bound states is
obtained by poles of $\mathcal{T}_\mathrm{o}(E)$.  When $E$ approaches
one of the binding energies $E_3<-\frac{\theta(a_\eff)}{2m_{AB}a_\eff^2}$,
we can write $\mathcal{T}_\mathrm{o}(E)$ as
$\mathcal{T}_\mathrm{o}(E;p_z,k_z)\to\mathcal{Z}_\mathrm{o}(p_z,q_z)/(E+|E_3|)$.
By solving the homogeneous integral equation from
Eq.~(\ref{eq:odd-parity}) satisfied by $\mathcal{Z}_\mathrm{o}(p_z,k_z)$
at the two-body resonance $|a_\eff|\to\infty$, we can obtain an infinite
series of binding energies $E_3^{(n)}$ expressed by the form of
Eq.~(\ref{eq:efimov}).  The ultraviolet cutoff is found to be related
with the Efimov parameter by $\Lambda=0.460\kappa_*$ 
for the mass ratio $m_A/m_B=6.67$ corresponding to the
${}^{40}$K-${}^6$Li mixture.  From now on we shall concentrate on this
most important case of $A={}^{40}$K and $B={}^6$Li.

Away from the two-body resonance $|a_\eff|<\infty$, there can be a
series of resonances associated with the Efimov trimers.  On the
positive side of the effective scattering length $a_\eff^{-1}>0$, the
three-body binding energy $E_3^{(n)}$ for a given $n$ decreases by
increasing the value of $a_\eff^{-1}$.  Eventually $E_3^{(n)}$ merges
into the atom-dimer threshold $E_3=-\frac1{2m_{AB}a_\eff^2}$ at the
critical effective scattering length given by
$a_\eff\kappa_*=0.0199\,e^{n\pi/s_0}$.
At such values of $a_\eff$, resonant behaviors in the atom-dimer
scattering are expected to occur~\cite{atom-dimer}.

Similarly, on the negative side of the effective scattering length
$a_\eff^{-1}<0$, $E_3^{(n)}$ increases by decreasing the value of
$a_\eff^{-1}$.  Eventually $E_3^{(n)}$ merges into the three-atom
threshold $E_3=0$ at the critical effective scattering length given by
$a_\eff\kappa_*=-1.89\,e^{n\pi/s_0}$. 
The three-body resonances at such values of $a_\eff$ shall be referred
to as confinement-induced Efimov resonances and bring significant
consequences on the three-body recombination rate for $a_\eff<0$.

\section{Three-body recombination rate}
The three-body recombination is an inelastic scattering process in which
two of three colliding atoms bind to form a diatomic molecule
($A{+}A{+}B{\to}A{+}AB$).  Assuming the binding energy of the dimer is
large enough so that the recoiling atom and dimer escape from the
system, the three-body recombination rate can be measured experimentally
through the particle loss rate of $A$ atoms:
$\dot{n}_A=-2\alpha n_A^2n_B$.  Here $n_{A(B)}$ is the
one-(three-)dimensional density of $A(B)$ atoms, and $\alpha$ is the
three-body recombination rate constant.  The other three-body
recombination channel ($A{+}B{+}B{\to}AB{+}B$), which does not exhibit
the Efimov effect as far as $m_A/m_B>0.00646$~\cite{Nishida:2008kr}, can
also contribute to the particle loss of $A$ atoms.  However, it is
subleading in the dilute limit $|a_\eff|\to0$ we consider below and thus
negligible.

A convenient way to compute $\alpha$ is to use the optical theorem which
relates $\alpha$ to twice the imaginary part of the forward three-body
scattering amplitude (Fig.~\ref{fig:3-body}).  In particular, in the
dilute limit where $|a_\eff|n_A\ll1$ and $|a_\eff|n_B^{1/3}\ll1$, the
odd-parity channel dominates the three-body scattering, and $\alpha$ can
be expressed as $\alpha=4\left(2\pi a_\eff/m_B\right)^2\mathrm{Im}\,\mathcal{T}_\mathrm{o}(0;p_z,p_z)\big|_{p_z\to0}$.
Because we can find
$\mathrm{Im}\,\mathcal{T}_\mathrm{o}(0,p_z,q_z)\big|_{p_z,q_z\to0}\propto p_z q_z$
from Eq.~(\ref{eq:odd-parity}), it is useful to introduce a
dimensionless function $t(q_z)$ by
$\mathcal{T}_\mathrm{o}(0;p_z,q_z)\big|_{p_z\to0}\equiv m_{AB}p_zq_za_\eff^2t(q_z)/\pi$.
Accordingly, the rate constant to the leading order in $a_\eff$ becomes
$\alpha=16\pi\left(m_{AB}/m_B^2\right)\bar p_z^2a_\eff^4\,\mathrm{Im}\,t(0)$.
Here $\bar p_z^2$ is the statistical average of the $A$ atom's momentum
squared and equals to $(\pi n_A)^2/3$ at zero temperature and $m_AT$ at
high temperature.

\begin{figure}[tp]
 \includegraphics[width=0.45\textwidth,clip]{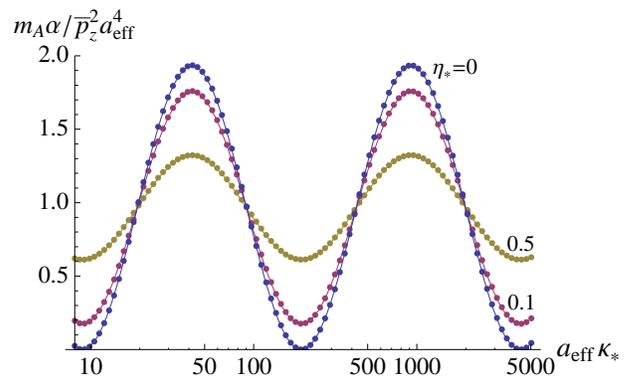}
 \caption{Two periods of $\frac{m_A}{\bar p_z^2a_\eff^4}\alpha$ as a
 function of $a_\eff\kappa_*>0$ at $\eta_*=0$, $0.1$, and $0.5$ for
 $m_A/m_B=6.67$.  Curves behind data points are from the two-parameter
 fit by Eq.~(\ref{eq:positive}).  \label{fig:positive}}
\end{figure}

Now $t(q_z)$ satisfies the integral equation
\begin{equation}\label{eq:integral}
  t(q_z) = \frac1{a_\eff^2q_z^2}
  + \int_0^\Lambda\!\frac{dk_z}{2\pi}\frac{k_z}{q_z}
  \frac{t(k_z)\,K(0;k_z,q_z)}
  {\sqrt{\frac{m_B+m_{AB}}{M}}\,k_z-\frac1{a_\eff}-i0^+}.
\end{equation}
It is clear that $t(0)$ has a nonzero imaginary part only for
$a_\eff>0$ in which the three-body recombination into the shallow dimer
is possible.  Remarkably the integral equation for $t(q_z)$ is quite
similar to that for the $s$-wave scattering amplitude of three identical
bosons in three dimensions~\cite{Bedaque:2000ft}, and therefore, the
solution has the similar property: $\mathrm{Im}\,t(0)$ is a
logarithmic-periodic function of $a_\eff\kappa_*$ with a scaling factor
$e^{\pi/s_0}=22.0$. 
The rate constant $\alpha$ for $a_\eff>0$ is plotted in
Fig.~\ref{fig:positive}.  We can see that the dimensionless quantity
$m_A\alpha/(\bar p_z^2a_\eff^4)$ oscillates between zero at
$a_\eff\kappa_*=0.404\,e^{n\pi/s_0}$ 
and the maximal value $1.93$ 
at $a_\eff\kappa_*= 1.89\,e^{n\pi/s_0}$. 
Such zeros in $\alpha$ have
been explained by the destructive interference effect between two decay
pathways in the case of identical bosons~\cite{Nielsen:1999}.

The effect of deeply bound dimers on the three-body recombination rate
can be taken into account by analytically continuing the Efimov
parameter to a complex value as
$\kappa_*\to e^{i\eta_*/s_0}\kappa_*$~\cite{Braaten:2004rn}.  Here
$\eta_*$ is a real positive parameter to make the Efimov trimers acquire
nonzero widths due to decays into the deeply bound dimers.  $\alpha$ for
$a_\eff>0$ at $\eta_*=0.1$ and $0.5$ are plotted in
Fig.~\ref{fig:positive} as well as at $\eta_*=0$.  We find that our
numerical solutions can be excellently fitted by the following formula
motivated by that for identical bosons~\cite{Braaten:2004rn}:
\begin{equation}\label{eq:positive}
 \begin{split}
  \frac{m_A}{\bar p_z^2a_\eff^4}\alpha
  &= b_+\frac{\sin^2[s_0\ln(c_+a_\eff\kappa_*)]+\sinh^2[\eta_*]}
  {\sinh^2[\pi s_0+\eta_*]+\cos^2[s_0\ln(c_+a_\eff\kappa_*)]} \\
  & + \frac{b_+}2\frac{\coth[\pi s_0]\sinh[2\eta_*]}
  {\sinh^2[\pi s_0+\eta_*]+\cos^2[s_0\ln(c_+a_\eff\kappa_*)]}.
 \end{split}
\end{equation}
Here $b_+=284$ 
and $c_+=2.48$ 
are two fitting parameters.

\begin{figure}[tp]
 \includegraphics[width=0.452\textwidth,clip]{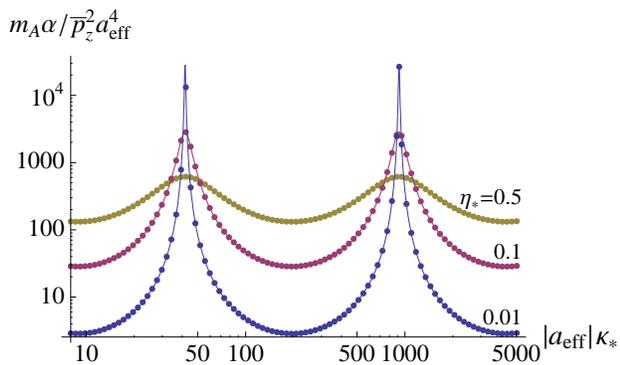}
 \caption{Two periods of $\frac{m_A}{\bar p_z^2a_\eff^4}\alpha$ as a
 function of $a_\eff\kappa_*<0$ at $\eta_*=0.01$, $0.1$, and $0.5$ for
 $m_A/m_B=6.67$.  Curves behind data points are from the two-parameter
 fit by Eq.~(\ref{eq:negative}).  \label{fig:negative}}
\end{figure}

When $\eta_*>0$, the solution to Eq.~(\ref{eq:integral}) can have a
nonzero imaginary part even for $a_\eff<0$ because the three-body
recombination into the deeply bound dimers becomes possible.  The rate
constant $\alpha$ for $a_\eff<0$ at $\eta_*=0.01$, $0.1$, and $0.5$ are
plotted in Fig.~\ref{fig:negative}.  Again we find that our numerical
solutions can be excellently fitted by the following
formula~\cite{Braaten:2004rn}:
\begin{equation}\label{eq:negative}
 \frac{m_A}{\bar p_z^2a_\eff^4}\alpha
  = b_-\frac{\sinh[2\eta_*]}
  {\sin^2[s_0\ln(c_-|a_\eff|\kappa_*)]+\sinh^2[\eta_*]},
\end{equation}
with two fitting parameters $b_-=143$ 
and $c_-=0.528$. 
We can see that when $\eta_*\ll1$, the three-body
recombination rate exhibits sharp resonant peaks at
$a_\eff\kappa_*=-1.89\,e^{n\pi/s_0}$, which are clear signatures of the
confinement-induced Efimov resonances.

Unlike the Efimov parameter $\kappa_*$, we cannot determine the width
parameter $\eta_*$ because it involves the complicated short-range
physics, but we can estimate $\eta_*$ to be very small in our system.
The size of the Efimov trimers is typically given by the confinement
length $l$.  In order for the Efimov trimer to decay into the deeply
bound dimers, the three bound atoms have to come within the range of an
interatomic potential $r_0(<l)$.  Its probability $\sim(r_0/l)^{4.39}$
for $m_A/m_B=6.67$~\cite{exponent} multiplied by the typical energy
scale of the short-range physics $\sim r_0^{-2}$ provides the
order-of-magnitude estimate of the width of the Efimov trimer.  The
width parameter is therefore found to be $\eta_*\sim(r_0/l)^{2.39}\ll1$
and scales with respect to $l$.  The small value of $\eta_*$ sharpens
the characteristic features in the three-body recombination rate such
as the destructive interferences at $a_\eff>0$ and the resonant peaks at
$a_\eff<0$ as seen in Figs.~\ref{fig:positive} and \ref{fig:negative}.

Finally we note that all the qualitative arguments presented in this
Rapid Communication hold for the ${}^{40}$K-${}^6$Li mixture when
${}^{40}$K is confined in 2D.  However, the scaling factor in the 2D-3D
mixed dimensions is $e^{\pi/s_0}=1.78\times10^5$, 
and therefore it is possible that the confinement-induced Efimov
resonances may not be observed in a range of the effective scattering
length surveyed by experiments.

\section{Conclusions}
We have shown that the Fermi-Fermi mixture of ${}^{40}$K and ${}^6$Li in
the 1D-3D mixed dimensions is a promising candidate to investigate the
Efimov physics in fermions.  Unlike other systems in a free space, the
Efimov parameter is controlled by the external confinement potential and
we can estimate the positions of the three-body resonances to be at
$a_\eff/l\approx-0.989\times(22.0)^n$. 
An observation of the confinement-induced Efimov resonances in the
three-body recombination rate at such predicted values of the effective
scattering length will provide the first unambiguous evidence of the
Efimov trimers in Fermi-Fermi mixtures.

\acknowledgments
Y.\,N.\ is supported by MIT Pappalardo Fellowship in Physics and
D.O.E. under cooperative research agreement DE-FG0205ER41360.
S.\,T.\ is supported by Yale University Postdoctoral Prize Fellowship.


\begin{thebibliography}{99}
\bibitem{error}
The Born-Oppenheimer approximation gives $0.907$ for $s_0$, while the
exact value is $1.02$.  Therefore the error of the Efimov parameter
computed here is expected to be of the order of 10\%.
\bibitem{atom-dimer}
However, in the low-energy limit, the odd-parity channel is subleading
in the atom-dimer scattering and the even-parity channel exhibiting no
Efimov effect is dominant.
\end{thebibliography}
\end{document}